\documentclass[aps,prd,twocolumn]{revtex4}

\usepackage{amssymb}
\usepackage{amsmath}

\usepackage{graphicx}
\usepackage[export]{adjustbox}

\usepackage{xparse}

\usepackage{hyperref}
\usepackage{hypernat}

\usepackage[caption=false,labelfont=normalsize,labelformat=empty,textfont=normalsize,justification=centering]{subfig}

\renewcommand{\Im}{\mathrm{Im}}

\newcommand{\iu}{\mathrm{i}}
\newcommand{\eq}{\mathrm{eq}}
\newcommand{\Fr}{\mathrm{Fr}}

\begin{document}

\title{CP Asymmetries and Higher-Order Unitarity Relations}

\author{Tom\'a\v s Bla\v zek}
\email{tomas.blazek@fmph.uniba.sk}
\author{Peter Mat\'ak}
\email{peter.matak@fmph.uniba.sk}
\affiliation{Department of Theoretical Physics, Comenius University,\\ Mlynsk\'a dolina, 84248 Bratislava, Slovak Republic}

\date{\today}

\begin{abstract}
The main focus of this paper is to introduce a new method to control perturbative calculations of CP asymmetric reaction rates in Boltzmann equation. CP asymmetries in particle reactions are traditionally calculated in terms of complex couplings, Feynman integrals, and Cutkosky rules. We use an expansion of the $S$-matrix unitarity condition instead, obtaining a general expression for the asymmetries without reference to the imaginary part of the loops. Asymmetry cancelations implied by CPT and unitarity are manifested in a diagrammatic way and easy to track at any order of perturbation theory. We demonstrate the power of this general framework within the right-handed neutrino and top-quark scattering asymmetries in seesaw type-I leptogenesis.
\end{abstract}

\maketitle

\section{INTRODUCTION.}
In physics, cancelations among terms in calculations may occasionally lead to small numbers or exact zero results that do not show up until the calculations' last steps are done. If this happens it may be suggestive that the language we use -- \emph{the way we cut reality into pieces} \cite{Kvasz:2015} --  is not fitting the description of the phenomenon. We argue that this may be the case of the $S$-matrix unitarity constraints relevant for contributions to particle-antiparticle asymmetries in the early universe. Such processes may include asymmetric decays of heavy particles  \cite{Kolb:1979qa,Weinberg:1979bt,Fukugita:1986hr}, scatterings \cite{Bento:2001rc,Yoshimura:1978ex}, annihilations \cite{Falkowski:2011xh} or multiparticle ($3\leftrightarrow 2$) reactions \cite{Nardi:2007jp}. They appear in different phenomenological contexts, such as baryogenesis \cite{Weinberg:1979bt,Kolb:1979qa}, leptogenesis \cite{Fukugita:1986hr, Bento:2001rc} or asymmetric dark matter models \cite{Kaplan:2009ag}. Interestingly enough, often the processes containing different particles, or even the processes of a different types, conspire together to cancel each other's contribution avoiding asymmetry generation in thermal equilibrium due to unitarity and CPT invariance \cite{Sakharov:1967dj,Hook:2011tk,Baldes:2014gca,Racker:2018tzw}. 

In type-I seesaw leptogenesis \cite{Roulet:1997xa,Buchmuller:1997yu,Flanz:1998kr,Frere:1999rh}, the asymmetry may come from heavy right-handed neutrino decays into lepton and Higgs boson, $N\leftrightarrow lH(\bar{l}\bar{H})$. However, the effect of the asymmetric $lH\leftrightarrow\bar{l}\bar{H}$ scattering has to be taken into account as well canceling the asymmetry in equilibrium. This is an example of what is known as a real intermediate state subtraction introduced in order to avoid double-counting in the Boltzmann collision term \cite{Kolb:1979qa}. 
Here our primary goal is to make such cancelations as evident as possible at any order of the perturbation theory. Moreover, we introduce a diagrammatic representation of asymmetries and reaction rates and formulate a general algorithm identifying the complete set of contributions.

We note that, to a certain extent, a similar approach has been used in \cite{Roulet:1997xa}. There, the contribution to the asymmetry of a reaction may be represented by a cyclic diagram cut into three pieces by \emph{in}, \emph{out} and \emph{Cutkosky} cuts. This, however, does not work that simple when any of the three parts contains on-shell intermediate particles or loop integrals with a non-zero imaginary part, or, in terms of the transition matrix introduced bellow when $T^*_{nm}\neq T_{mn}$ for any of the three parts. In contrast, our approach is fully general and can be applied both to real intermediate state subtractions and higher-order terms with complex loops.

\section{CP VIOLATION AND $S$-MATRIX EXPANSION.}
The $S$-matrix unitarity condition $S S^\dagger = S^\dagger S = 1$ written in terms of the transition matrix $T_{fi}=\iu(\delta_{fi}-S_{fi})=(2\pi)^4\delta^{(4)}(p_f-p_i)M_{fi}$ for the initial state $\vert i\rangle$ and the final state $\vert f\rangle$ leads to the relation
\begin{equation}\label{eq1}
T^\dagger_{fi}=T^{\vphantom{\dagger}}_{fi}-\iu\sum_n T^\dagger_{fn} T^{\vphantom{\dagger}}_{ni}.
\end{equation}
Then, within a CPT symmetric quantum theory, the CP asymmetry $\Delta \vert T_{fi}\vert^2=\vert T_{fi}\vert^2 - \vert T_{if}\vert^2$ can be written as \cite{Kolb:1979qa}
\begin{equation}\label{eq2}
\Delta \vert T^{\vphantom{\dagger}}_{fi}\vert^2=2 \Im\sum_n T^{\vphantom{\dagger}}_{in}T^\dagger_{nf}T^{\vphantom{\dagger}}_{fi}-\Big\vert\sum_n T^\dagger_{fn} T^{\vphantom{\dagger}}_{ni}\Big\vert^2.
\end{equation}
Now, let us replace each $T^\dagger$ in Eq. \eqref{eq2} using Eq. \eqref{eq1}. Repeating the procedure iteratively we obtain an expansion
\begin{eqnarray}\label{eq3}
\Delta \vert T^{\vphantom{\dagger}}_{fi}\vert^2 &=&\sum_{n}(\iu T_{in} \iu T_{nf} \iu T_{fi} - \iu T_{if} \iu T_{fn} \iu T_{ni})\\
&-&\sum_{n,m}(\iu T_{in} \iu T_{nm} \iu T_{mf} \iu T_{fi} - \iu T_{if} \iu T_{fm} \iu T_{mn} \iu T_{ni})\nonumber\\
&+&\vphantom{\sum_{n}}\ldots\,.\nonumber
\end{eqnarray}
Alternatively, the unitarity condition \eqref{eq1} may be used directly to derive 
\begin{equation}
\Delta \vert T^{\vphantom{\dagger}}_{fi}\vert^2=\iu T_{if}(T^2 S^\dagger)\vphantom{T}_{fi} - (S^\dagger T^2)\vphantom{T}_{if} \iu T_{fi}.
\end{equation}
This compact expression is equivalent to Eq. \eqref{eq3} by the expansion of $S^\dagger=S^{-1}$ as a geometric series. We can observe that, unlike in Eq. \eqref{eq2}, the cancelation of asymmetries in Eq. \eqref{eq3} after the summation over the final states is evident from the opposite signs of the mirrored terms in brackets. Moreover, there are no conjugated amplitudes. A single term in Eq. \eqref{eq3} may be viewed as a transition from $\vert i\rangle$ to $\vert i\rangle$ with two or more on-shell intermediate states including $\vert f\rangle$. To generate all the contributions of this type we introduce their diagrammatic representation. First, we write forward scattering amplitudes in terms of Feynman diagrams to a certain order in coupling constants. Then, we make all the cuts that are kinematically allowed (see Fig. \ref{fig1} bellow). Any subset of two or more cuts, with at least one vertex in each piece, will represent a contribution to the asymmetry. However, it is important to note that (unlike the Cutkosky cuts) here \emph{cutting} refers to simply putting certain particles on a mass shell, without changing the sign of the $\iu\epsilon$ in any propagator on any side of the cut.

\section{ASYMMETRIES IN BOLTZMANN COLLISION INTEGRAL.}
In ma\-ny scenarios of matter generation in the ex\-pan\-ding universe, the Boltzmann equation gives a reasonably accurate approximation to the final relic densities and asymmetries. Here we consider the particles interacting as if they appear in a vacuum, using the Maxwell-Boltzmann statistics. The generalization, including the correct statistical factors for the final states, can be found following the discussion in \cite{Hook:2011tk,Nanopoulos:1979gx,Blazek:2021zoj}. However, before we start with the kinetic description of particle interactions, we need to find an efficient way to deal with the products of transition matrices as they appear in Eq. \eqref{eq3}. There, the summation includes all the combinations of distinct particle species allowed by symmetries and kinematics. The momentum integration and the summations over the discrete degrees of freedom are included as well. From now on, we change this notation. The sum will only run over different particles in the intermediate states. The momentum integrals and spin summations will be indicated by the \emph{cut product} defined as
\begin{equation}\label{eq5}
\sum_{\forall s_n} \int\prod_{\forall p_n}[d\mathbf{p}_n] \iu T_{in} \iu T_{nf}\stackrel{\mathrm{def.}}{=}\iu T_{in}\vert\iu T_{nf}
\end{equation}
using $[d\mathbf{p}_n] = d^3\mathbf{p}_n/((2\pi)^3 2E_n)$ for the Lorentz invariant measure in the momentum space.

Now, let us consider the equilibrium contribution of the $i\rightarrow f$ reaction to the evolution of the number densities of the included particles. Within the classical Boltzmann approach, it is given in terms of the thermally averaged rate of the process
\begin{equation}\label{eq6}
\gamma^\eq_{fi}
=-\frac{1}{V_4}\sum_{\forall s_i} \int\bigg(\prod_{\forall p_i}[d\mathbf{p}_i]f^{\eq}_i(p_i)\bigg) \iu T^\dagger_{if}\vert\iu T^{\vphantom{\dagger}}_{fi},
\end{equation}
where $V_4$ denotes the four-dimensional volume \footnote{ Alternatively, we may define the product of amplitudes as
$(2\pi)^4\delta^{(4)}(p_f-p_i)\iu M_{fm}\vert\ldots\vert\iu M_{ni} = \iu T_{fm}\vert\ldots\vert\iu T_{ni}$. Then, the expressions rewritten in terms of $M$'s will be free of the infinite volume.}. Looking at Eq. \eqref{eq6}, we can observe its trace-like structure. Indeed, in equilibrium, cyclicity is guaranteed by the detailed balance condition \cite{Bernstein:1988bw}. Thus, we can define the trace of the product of transition matrices
\begin{widetext}
\begin{equation}\label{eq7}
\Fr\{\iu T_{in}\vert\iu T_{nm}\vert\ldots\vert\iu T_{fi}\}\stackrel{\mathrm{def.}}{=}\frac{1}{V_4}\sum_{\forall s_i} \int\bigg(\prod_{\forall i}[d\mathbf{p}_i]f^{\eq}_i(p_i)\bigg) \iu T_{in}\vert\iu T_{nm}\vert\ldots\vert\iu T_{fi},
\end{equation}
\end{widetext}
where we have changed $\mathrm{Tr}$ (trace) to a new symbol $\Fr$ emphasizing the presence of the equilibrium phase space densities (as we compute traces over the \emph{forest} of Feynman diagrams).
Using the expansion \eqref{eq3} we obtain for the equilibrium asymmetry of the $i\rightarrow f$ reaction rate
\begin{eqnarray}\label{eq8}
\Delta\gamma^\eq_{fi}&=&\sum_{n}\Fr\{\iu T_{in}\vert\iu T_{nf}\vert\iu T_{fi}\}-\mathrm{m.t.}\\
&-&\sum_{n,m}\Fr\{\iu T_{in}\vert\iu T_{nm}\vert\iu T_{mf}\vert\iu T_{fi}\}-\mathrm{m.t.}\nonumber\\
&+&\nonumber\ldots 
\end{eqnarray}
with m.t. representing the \emph{mirrored terms}, in which the intermediate states appear in reversed order.

As an elementary example, let us consider the asymmetric right-handed neutrino decays mentioned earlier. We use the Lagrangian density
\begin{equation}\label{eq9}
\mathcal{L} \supset -\frac{1}{2}M_i\bar{N}_i N_i - \Big(\mathcal{Y}_{\alpha i}\bar{N}_i P_L l_{\alpha} H + \mathrm{H.c.}\Big)
\end{equation}
where $i$ and $\alpha$ are family indices labeling the right-handed neutrino and standard model leptons, respectively. 
\begin{figure*}
\subfloat{\label{fig1a}}
\subfloat{\label{fig1b}}
\subfloat{\label{fig1c}}
\subfloat{\label{fig1d}}
\includegraphics{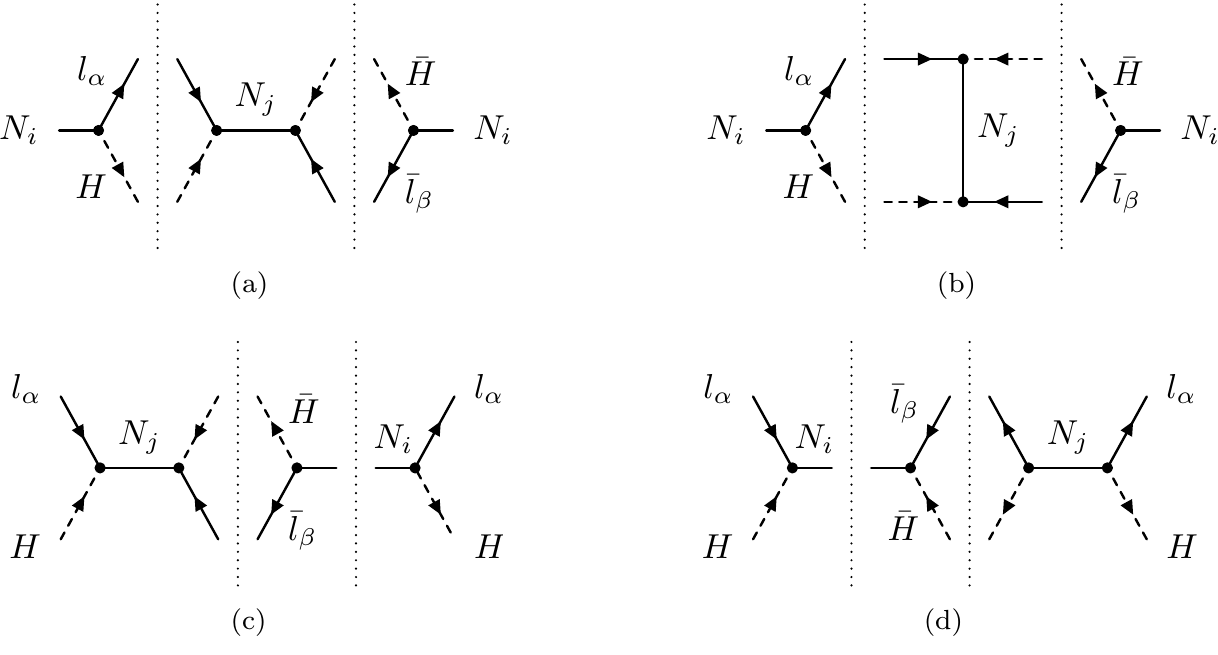}
\caption{\label{fig1} Lepton number violating contributions to $\Delta\vert T(N\rightarrow lH)\vert^2$ (Fig. \ref{fig1a}, \ref{fig1b}), $\Delta\vert T(lH\rightarrow\bar{l}\bar{H})\vert^2$ (Fig. \ref{fig1c}) and $\Delta\vert T(\bar{l}\bar{H}\rightarrow N)\vert^2$ (Fig. \ref{fig1d}). For the scattering and inverse decay, additional $t$-channel diagrams must be included as well.}
\end{figure*}
The diagrams depicted in Fig. \ref{fig1a} and \ref{fig1b} (minus the mirrored terms) represent the self-energy and vertex part of the $N\rightarrow lH$ decay asymmetry, respectively. On the other hand, the diagram in Fig. \ref{fig1c} enters the asymmetry of the $lH\rightarrow\bar{l}\bar{H}$ scattering. It is equivalent to the $s$-channel part of the real intermediate state subtracted scattering rate in the narrow width approximation \cite{Flanz:1998kr}. To see this more clearly, let us consider the right-handed neutrino propagator in the form of
\begin{equation}\label{eq10}
	\frac{\iu}{p^2-M^2_i+\iu\Gamma_i M_i}
\end{equation}
where $\Gamma_i$ represents the total width. Taking the limit $\Gamma_i/M_i\rightarrow 0$ then leads to 
\begin{equation}\label{eq11}
\mathrm{P.V.}\,\frac{\iu}{p^2-M^2_i}-\pi\delta(p^2-M^2_i)
\end{equation}
that, in the case of $lH\rightarrow\bar{l}\bar{H}$ scattering, can be diagrammatically represented as
\begin{widetext}
	\begin{equation}\label{eq12}
		\includegraphics[scale=1, valign=c]{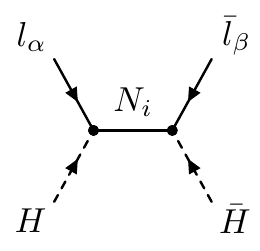}=\mathrm{P.V.}\includegraphics[scale=1, valign=c]{EQfig5a.pdf}+\frac{1}{2}\includegraphics[scale=1, valign=c]{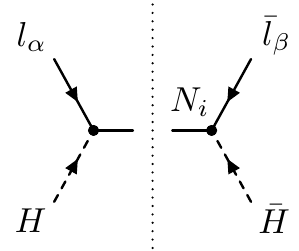}
	\end{equation}
\end{widetext}
where the cut over the right-handed neutrino line in the second term is understood as in Eq. \eqref{eq5}. Summing over the neutrino family $i$, taking the square of the above expression and subtracting analogous contribution of the CP conjugated reaction, leads exactly to what we obtain from Fig. \ref{fig1c} minus the mirrored terms in Fig. \ref{fig1d}. On the other hand, we can split Eq. \eqref{eq10} into the real and imaginary part and take the square. The term containing the square of the width in the numerator, in the narrow width limit, gives the real intermediate state that has to be subtracted \cite{Kolb:1979qa,Hook:2011tk}. From the cyclicity of the trace in Eq. \eqref{eq7} we can immediately see that these contributions (Fig. \ref{fig1c}, \ref{fig1d}) are equal to Fig. \ref{fig1a} minus its mirrored terms. The same, up to the sign, applies to Fig. \ref{fig1d} corresponding to the inverse decay process. 

Therefore, at this order in the neutrino Yukawa coupling, we only have two independent contributions entering the relevant processes' asymmetries - those containing the $s$- or $t$-channel neutrino propagator in the scattering diagram depicted in Fig. \ref{fig1a} and \ref{fig1b}. Everything else may be expressed in terms of their cyclic permutations and conjugations.

\section{NEUTRINO-QUARK SCATTERING AND $\mathcal{O}(\mathcal{Y}^4\mathcal{Y}^2_t)$ UNITARITY RELATIONS.}
The lowest order asymmetries are simple, even within the standard Cutkosky approach. Here, as a more advanced example, we consider the scattering of right-handed neutrino and top-quark \cite{Racker:2018tzw,Nardi:2007jp,Pilaftsis:2003gt}. To this purpose, we add to the Lagrangian density \eqref{eq9}
\begin{equation}\label{eq13}
\mathcal{L} \supset -\mathcal{Y}_t\bar{t} P_L Q H + \mathrm{H.c.},
\end{equation}
with $t$ and $Q$ representing the right-handed top and left-handed quark doublet, respectively. All particles, except for the right-handed neutrinos, are considered massless. As we already mentioned, given the initial state $N_i Q$ we should take all the corresponding forward scattering diagrams and cut them into three, four, or as many pieces as possible. What particular vacuum diagrams should we start with?
\begin{figure}
\centering
\includegraphics{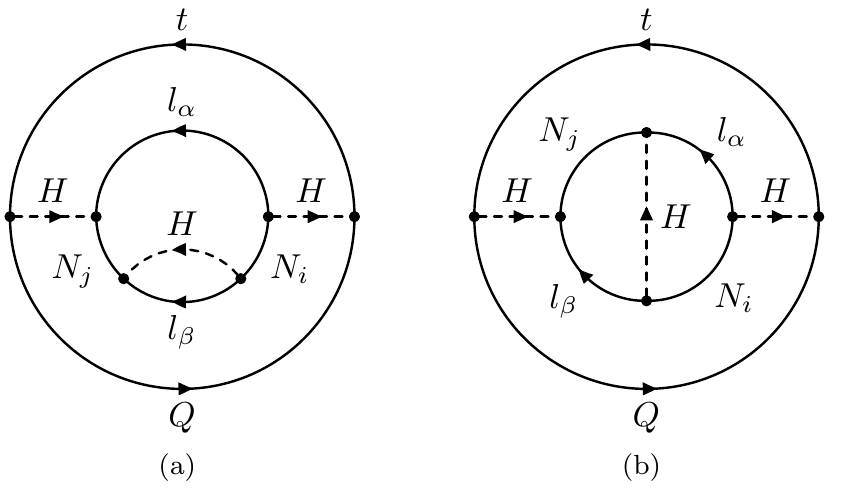}
\caption{\label{fig2} Lepton number violating vacuum diagrams leading to asymmetries at the $\mathcal{O}(\mathcal{Y}^4\mathcal{Y}^2_t)$ order. Similar vacuum diagrams were introduced in \cite{Racker:2018tzw} and, in a different context in \cite{Botella:2004ks}, where the standard Cutkosky approach has been used.}
\end{figure}
To obtain a reliable answer, we start with all vacuum bubbles made out of four neutrino and two top-quark Yukawa vertices. There are six of them. One comes with $\vert\mathcal{Y}_{\alpha i}\vert^2\vert\mathcal{Y}_{\beta j}\vert^2\mathcal{Y}^2_t$, a real combination of couplings that does not affect the asymmetry. Three bubbles can be made in a lepton number conserving way, proportional to $\mathcal{Y}^*_{\alpha i}\mathcal{Y}^{\vphantom{*}}_{\alpha j}\mathcal{Y}^{\vphantom{*}}_{\beta i}\mathcal{Y}^*_{\beta j}\mathcal{Y}^2_t$ . Their contribution vanishes after summing over the lepton flavors. However, it is important for the complete description of the lepton number generation \cite{Abada:2006fw,Nardi:2006fx,AristizabalSierra:2009bh,Abada:2006ea}. This  is not the purpose of the present work. To illustrate our method, we consider the lepton number violating vacuum diagrams as shown in Fig. \ref{fig2}. There is one $Q$ propagator in each of them and one way of cutting it to obtain the $N_i Q \rightarrow N_i Q$ diagram. There are two right-handed neutrino propagators and two ways of cutting each of them. The resulting diagrams are shown in Fig. \ref{fig3}, where some of the family indices have been relabeled so that $i$ corresponds to the initial right-handed neutrino while $\alpha$ and $\beta$ label the lepton and antilepton, respectively \footnote{The vacuum diagrams that we started from may remind us of thermal field theory with two-particle irreducible effective action. Here, however, the (reducible) bubble diagrams are only used to ensure we generate all forward scattering diagrams without forgetting anything. The connection of the present work to the field theory at finite temperature is investigated in a separate study.}. To generate the $T$'s of Eq. \eqref{eq8} we consider all possible cuts of these diagrams into three or more pieces. Each piece stands for one amplitude $T$. Note that $T_{fi}$, the first piece of each diagram, defines the process.
\begin{figure*}
\subfloat{\label{fig3a}}
\subfloat{\label{fig3b}}
\subfloat{\label{fig3c}}
\subfloat{\label{fig3d}}
\subfloat{\label{fig3e}}
\subfloat{\label{fig3f}}
\subfloat{\label{fig3g}}
\subfloat{\label{fig3h}}
\includegraphics{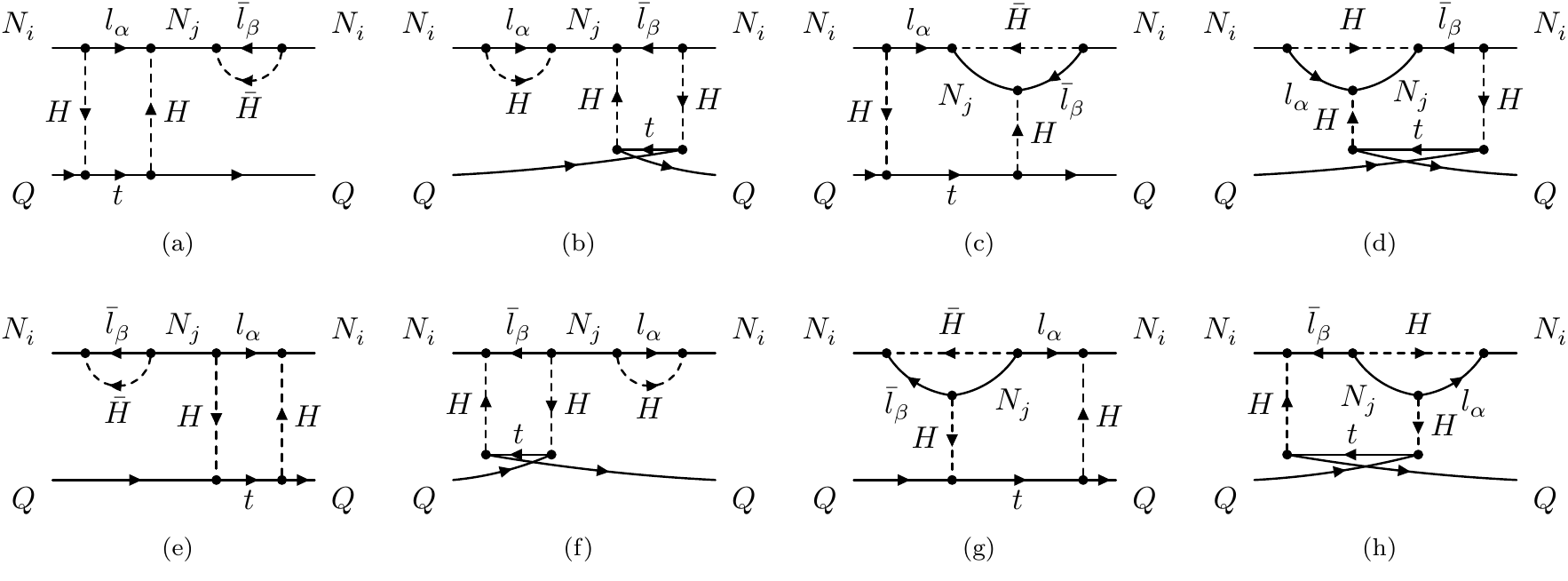}
\caption{\label{fig3} Forward scattering diagrams for the $N_i Q$ initial state obtained from Fig. \ref{fig2}. Cutting them will generate asymmetry relations for reactions with two-particle color triplet initial states at the $\mathcal{O}(\mathcal{Y}^4\mathcal{Y}^2_t)$ order.}
\end{figure*}

We are now ready to consider the relevant ways of cutting the diagram Fig. \ref{fig3a}. Four different cuts can be made, but only a certain combination of these can affect the asymmetry. If, for example, the $\bar{l}_\beta\bar{H}$ loop on the right-handed neutrino leg remains uncut, there is only one way to split this diagram into three pieces (see Eq. \eqref{eq8}). It requires two cuts of the $l_\alpha$ line, each cutting one of the Higgs box propagators. The corresponding mirrored terms add to the asymmetry of the same process, $N_i Q\rightarrow lHQ$. However, they are completely canceled out by cuts of the diagram in Fig. \ref{fig3e}. If, on the other hand, the $\bar{l}_\beta\bar{H}$ loop is cut, we get contributions to the asymmetries of $N_i Q\rightarrow l t$, $N_i Q\rightarrow lHQ$ and, with a minus sign, to the mirrored terms of $N_i Q\rightarrow \bar{l}\bar{H}Q$. Now, cutting the diagram in Fig. \ref{fig3e} will add to the asymmetry of $N_i Q\rightarrow \bar{l}\bar{H}Q$ and the mirrored terms of $N_i Q\rightarrow lt$, $N_i Q\rightarrow lHQ$. Using the expansion of the reaction rate asymmetry in Eq. \eqref{eq8}, we can write
\begin{widetext}
\begin{subequations}\label{eq14}
\begin{alignat}{1}
\Delta\gamma^{(a)}_{N_i Q\rightarrow lt} &= \Fr\includegraphics[scale=1, valign=c]{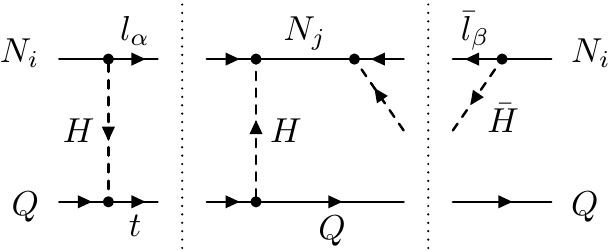}-\mathrm{m.t.},\label{eq14a}\\
\Delta\gamma^{(a)}_{N_i Q\rightarrow lHQ} &= \Fr\includegraphics[scale=1, valign=c]{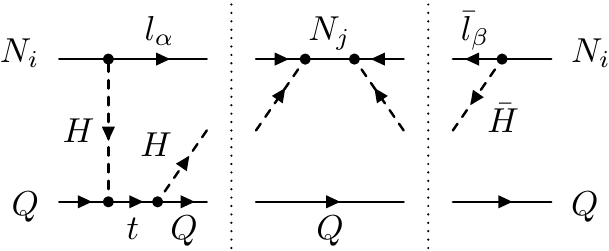}+\Fr\includegraphics[scale=1, valign=c]{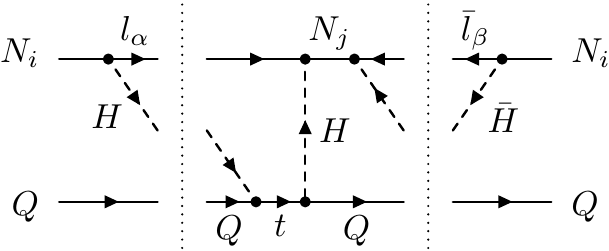}\label{eq14b}\\
&-\Fr\includegraphics[valign=c]{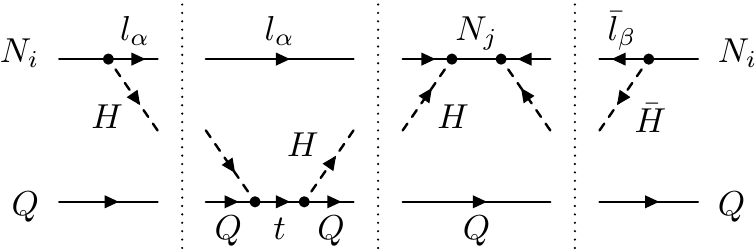}-\mathrm{m.t.},\nonumber\\
\Delta\gamma^{(a)}_{N_i Q\rightarrow\bar{l}\bar{H}Q} &= -\Delta\gamma^{(a)}_{N_i Q\rightarrow lt}-\Delta\gamma^{(a)}_{N_i Q\rightarrow lHQ}.\label{eq14c}
\end{alignat}
\end{subequations} 
\end{widetext}
Repeating the procedure pairwise for all the diagrams in the upper and lower rows of Fig. \ref{fig3} (and sorting them out according to reactions, namely by the first piece) leads straightforwardly to the $\mathcal{O}(\mathcal{Y}^4\mathcal{Y}^2_t)$ unitarity relation
\begin{eqnarray}\label{eq15}
0&=&\Delta\gamma^\eq_{N_iQ\rightarrow lt}+\Delta\gamma^\eq_{N_iQ\rightarrow lHQ}\\
&+&\Delta\gamma^\eq_{N_iQ\rightarrow\bar{l}\bar{H}Q}+\Delta\gamma^\eq_{N_iQ\rightarrow\bar{l}QQ\bar{t}}\,.\nonumber
\end{eqnarray}
For the sake of simplicity, the processes occurring above the $M_j$ threshold have been neglected in Eq. \eqref{eq15}. Analogous relations may be written for the $N_i\bar{t}$ initial state using the apparent $Q\leftrightarrow \bar{t}$ symmetry of Fig. \ref{fig2} or, using Eq. \eqref{eq3}, for unaveraged asymmetries $\Delta\vert T_{fi}\vert^2$ instead of thermally averaged $\Delta\gamma^\mathrm{eq}_{fi}$. Moreover, in Eq. \eqref{eq15}, the novel and interesting part is the presence of $2\leftrightarrow 4$ reaction that is related to the forward diagrams in Fig. \ref{fig3b}, \ref{fig3d}, \ref{fig3f} and \ref{fig3h}. To our knowledge, they were not taken into account in previous works, even though considering $2\leftrightarrow 3$ reactions, within the classical Boltzmann approach \cite{Nardi:2007jp,Racker:2018tzw,Abada:2006ea,Pilaftsis:2003gt,Pilaftsis:2005rv}.

\section{EVALUATION OF ASYMMETRIES AND INFRARED FINITENESS.}

In this section, we briefly discuss technical aspects of the diagrammatic representation used in Eq. \eqref{eq14} and comment on their relation to previous works. 

Splitting the uncut Higgs propagators in Eq. \eqref{eq14b} into the principal value and Dirac delta function, along the same lines as in Eqs. \eqref{eq11} and \eqref{eq12}, cancels the last singular term with three cuts. This feature, however, is not generic, and there are examples of diagrams where (non-singular) higher-order terms of the expansion \eqref{eq8} persist \footnote{For example, the scattering of $l\phi_a$ within the model considered in Ref. \cite{Kayser:2010fc} leads to diagrams of this type.}. The remaining terms in Eq. \eqref{eq14b} no longer contain $\iu\epsilon$ in propagators, and the resulting asymmetry is proportional to the couplings' imaginary part. The particular example of cutting the diagram in Fig. \ref{fig3a} can be evaluated in a similar way as the CP symmetric part of the contribution to the $N_iQ\rightarrow lt$, $N_i Q\rightarrow lHQ$ reactions that has been discussed in Ref. \cite{Racker:2018tzw} using the procedure suggested in Ref. \cite{Frye:2018xjj}. In the first two terms in Eq. \eqref{eq14b}, the principal value of the uncut Higgs propagator appears together with the delta function due to the on-shell Higgs. Then the integration over the Higgs momenta is carried out using the identity \cite{Frye:2018xjj,Racker:2018tzw}
\begin{equation}\label{eq16}
2\theta(p^0)\delta(p^2)\mathrm{P.V.}\frac{1}{p^2} = -\frac{1}{(p^0+\vert\mathbf{p}\vert)^2}\frac{\partial\delta(p^0-\vert\mathbf{p}\vert)}{\partial p^0}.
\end{equation}
The result contains infrared divergences that can be regulated by non-zero quark masses. They are canceled by terms resulting form Eq. \eqref{eq14a}.
Ref. \cite{Racker:2018tzw} observes this cancelation for the symmetric part of the reaction rate. 

All diagrams in Fig. \ref{fig3} can be cut in a way analogous to Eq. \ref{eq14}. However, the diagrams in Fig. \ref{fig3b}, \ref{fig3d}, \ref{fig3f} and \ref{fig3h} are special. The central cut through $\bar{l}$, $\bar{t}$, and two $Q$ lines is only kinematicaly allowed if
\begin{equation}\label{eq17}
	s=(p_{N_i}+p_Q)^2<2M^2_i.
\end{equation}
The use of particle description (instead of the formalism of non-equilibrium quantum field theory) is only justified for temperature that is low compared to the right-handed neutrino masses. If this condition is met, the remaining contributions of these diagrams can be multiplied by $\theta(2M^2_i-s)$ before the thermal average is performed. Then the finite part of the reaction rate is only affected mildly, while the Kinoshita-Lee-Nauenberg theorem guarantees the infrared finiteness. Above the threshold in Eq. \eqref{eq17}, the forward $N_iQ\rightarrow N_iQ$ scattering is needed to cancel the infrared divergences \cite{Frye:2018xjj}. This, however, is difficult to include in the Boltzmann equation and calls for proper treatment of the thermal effects, which is beyond the scope of the present paper. It will be presented in separate work in the future. 

Let us finally present how the asymmetries in Eq. \eqref{eq14} may be evaluated using the standard Cutkosky rules. Dividing the tree- and loop-level parts of the $i\rightarrow f$ amplitude into couplings ($C$) and kinematics ($K$), the resulting asymmetry can be expressed as
\begin{equation}\label{eq18}
	\Delta\vert T_{fi}\vert^2=-4\Im\left[C^{\vphantom{*}}_{\mathrm{tree}}C^*_{\mathrm{loop}}\right]\Im\left[K^{\vphantom{*}}_{\mathrm{tree}}K^*_{\mathrm{loop}}\right].
\end{equation}
Usually, the tree-level part is real, and one can write
\begin{equation}\label{eq19}
	\Im\left[K^{\vphantom{*}}_{\mathrm{tree}}K^*_{\mathrm{loop}}\right]=\frac{\iu}{2}K^{\vphantom{*}}_{\mathrm{tree}}\sum_{\mathrm{cuts}}K^{\vphantom{*}}_{\mathrm{loop}}.
\end{equation}
However, this formula can not be used in the case of $N_i Q\rightarrow lHQ$ reaction in Eq. \eqref{eq14b}, in which the tree-level diagram
\begin{equation}\label{eq20}
	\includegraphics[scale=1, valign=c]{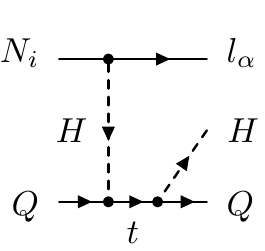}
\end{equation}
comes with the Higgs propagator containing the on-shell imaginary part from
\begin{equation}\label{eq21}
	\frac{1}{p^2+\iu\epsilon}=\mathrm{P.V.}\frac{1}{p^2}-\iu\pi\delta(p^2).
\end{equation}
Nevertheless, Eq. \eqref{eq18} remains valid and can be used to extract the asymmetry directly from the real and imaginary part of the corresponding loop diagram. The second term on the right-hand side of Eq. \eqref{eq21} then leads to a singular part containing $\delta(p^2)^2$ with the second delta function coming from the on-shell Higgs in the final state, as it is in Eq. \eqref{eq16}. This is canceled by an analog of the second term in Eq. \eqref{eq14b}. Our approach, based on the expansion of Eq. \eqref{eq3}, may be understood as a modification of Eq. \eqref{eq19}, in which these cancelations become visible at the diagrammatic level of Eq. \eqref{eq14}.

\section{Summary.}
Using the expansion of the $S$-matrix unitarity condition we have obtained a diagrammatic representation of the CP asymmetries in $\Delta\vert T_{fi}\vert^2$ and thermally averaged reaction rates $\Delta\gamma^\mathrm{eq}_{fi}$. We generate a complete set of contributions to the reaction's asymmetries starting with cuts of the vacuum diagrams at particular order in coupling constant. Unlike the standard Cutkosky approach that uses a single cut to determine the loop's imaginary part we introduce  multiple cuts of the corresponding forward scattering diagrams. With this procedure the cancelations among the resulting asymmetries are easy to track. Our general framework was demonstrated in an  analysis of the higher-order corrections to the right-handed neutrino scatterings in seesaw type-I leptogenesis.

\begin{acknowledgements}
We would like to thank our colleague, Vladim\'ir Balek, for useful comments and discussion. The authors were supported by the Slovak Ministry of Education five-year contract 0211/2016.
\end{acknowledgements}

\bibliographystyle{apsrev4-1.bst}
\bibliography{CLANOK.bib}

\end{document}